\documentclass[11pt]{cernrep}
\usepackage{graphicx}
\usepackage{here}
\begin{document}
 \title{Jet and Dijet Rates in AB Collisions}
\author{A. Accardi$^a$, N. Armesto$^b$ and I. P. Lokhtin$^c$}
%\author{A. Accardi$^a$, N. Armesto$^b$, I. P. Lokhtin$^c$
%and U. A. Wiedemann$^d$}
\institute{$^a$ Institut f\"ur Theoretische Physik der Universit\"at
Heidelberg, Germany\\
$^b$ Departamento de F\'{\i}sica, Universidad de C\'ordoba, Spain\\
$^c$ Institute of Nuclear Physics, Moscow State University, Russia
%\\$^d$ Theory Division, CERN, Switzerland
}
\maketitle
%\begin{abstract}
%\end{abstract}

Jet studies will play a central role as a proposed signature of the
formation of QGP
in AB collisions. Energy loss of energetic partons inside a medium
where colour charges are present, the
so-called jet quenching \cite{quench}, has been suggested to behave very
differently in cold nuclear matter and in QGP, and postulated as a tool to
probe the properties of this new state of matter \cite{tomo}.

On the other hand, jet calculations at NLO have been successfully
confronted with experimental data in hadron-hadron
collisions \cite{cdfjets}. Monte Carlo codes have become available:
among them, we will use that of \cite{nlocode,nlocode2} adapted to
include isospin effects and modifications of nucleon pdf inside nuclei,
see the Section
on Jet and Dijet Rates in pA Collisions \cite{pAcoll} for more information.
Here we will present the results of 'initial' state effects, i.e. no
energy loss of any kind will be considered. These results can be
considered as the reference, hopefully to be tested in pA, whose failure
should indicate the presence of new physics.
As in pA collisions,
we will work in the LHC lab frame, which for symmetric AB collisions
coincide with the center-of-mass one, and the accuracy of our
computations, limited by CPU time, is the same as in the pA case.
 
Unless explicitly stated and as in the pA case,
we will use as nucleon pdf MRST98 central
gluon \cite{mrst98} modified inside nuclei using the EKS98
parameterizations \cite{eks98}, a factorization scale equal to the
renormalization scale $\mu=\mu_F=\mu_R=E_T/2$, with $E_T$ the total
transverse energy of all the jets in the generated event,
and for jet reconstruction we will employ
the $k_T$-clustering algorithm \cite{ktal} with $D=1$.
The kinematical regions we
are going to consider are the same as in the pA case:
\begin{itemize}
\item $|\eta_i|<2.5$, with $\eta_i$ the pseudorapidity of the jet; this
corresponds to
the acceptance of the central part of the CMS detector.
\item $E_{Ti}>20$ GeV in the pseudorapidity distributions, with
$E_{Ti}$ the transverse energy of the jet; this will ensure the validity
of perturbative QCD.
\item $E_{T1}>20$ GeV and $E_{T2}>15$ GeV for the $\phi$-distributions,
with $E_{T1}$ ($E_{T2}$) the transverse energy of the hardest
(next-to-hardest) jet entering the CMS acceptance, and $\phi$ the angle
between these two jets.
\end{itemize}
Please have a look to the mentioned Section on pA to obtain more
information. As we did there, no centrality dependence is studied in
this article.

The words of caution about our results which were given in the pA
Section are even more relevant in AB collisions, as our ignorance on soft
multiparticle production in this case is even larger than in pA
collisions. For
example, the number of particles produced at midrapidity in a central
PbPb collision at the LHC may vary as much as a factor 3
\cite{simulators}
among different
models which, in principle, are able to reproduce the available
experimental data on multiplicities at SPS, RHIC and TeVatron.
Therefore, these
issues of the underlying event \cite{under}
and multiple hard parton scattering \cite{mhps,cdfmhps}
demand extensive Monte Carlo studies including full detector simulation.
Preliminary analysis, based on the developed sliding window-type jet
finding algorithm (which subtracts the large background from the
underlying event) and full GEANT-based simulation of the CMS
calorimetry,
shows that even in the worst case of central PbPb collisions with
maximal
estimated charged particle density at mid-rapidity $dN^{\pm}/dy|_{y=0}=8000$,
jets can be reconstructed with almost 100~\% efficiency, low noise and
satisfactory energy and spatial resolution starting from $E_{Ti} \sim 100$
GeV (see the Section on Jet Detection at CMS).
In the case of more realistic, lower multiplicities,
the
minimal threshold for adequate jet reconstruction could even decrease.
% strikman's comments

As in the pA case, see the previously mentioned Section on pA collisions
\cite{pAcoll}, the influence of disconnected collisions on jet
production in AB collisions may be studied using
simple estimates in AB collisions on the number $\langle n
\rangle$ of
nucleon-nucleon collisions involved in the production of jets with
$E_{Ti}$ greater than a given $E_{T0}$, which
can be obtained in the Glauber
model \cite{ccrit} in the optical approximation: $\langle n \rangle
(b,E_{T0})=ABT_{AB}(b)\sigma(E_{T0})/\sigma_{AB}(b,E_{T0})$, with $b$ the
impact parameter, $T_{AB}(b)=\int d^2s T_{A}(s)T_{B}(b-s)$
the convolution of the nuclear profile functions of projectile and
target normalized to unity, $\sigma(E_{T0})$ the
cross section for production of jets with
$E_{Ti}$ greater than $E_{T0}$ in pp collisions, and
$\sigma_{AB}(b,E_{T0})=1-[1-T_{AB}(b)\sigma(E_{T0})]^{AB}$. Taking
$\sigma(E_{T0})=70$, 0.1
and 0.006 $\mu$b as representative values in PbPb collisions at 5.5 TeV
for $E_{T0}=20$, 100 and 200 GeV respectively (see results
in Fig.
\ref{abfig2-4} below), the number
of nucleon-nucleon collisions involved turns out to be respectively
1.6, 1.0 and 1.0 
for minimum bias collisions (i.e. integrating numerator and
denominator in $\sigma_{pA}(b,E_{T0})$ between $b=0$ and $\infty$),
while for central collisions (integrating between $b=0$ and 1 fm) the
numbers are 2.4, 1.0 and 1.0 respectively.
So, in AB collisions at LHC
energies the contribution of multiple hard scattering coming from
different nucleon-nucleon collisions seems to be negligible for
transverse energies of the jets greater than
$\sim 100$ GeV, while for $E_{Ti}$ smaller
than $\sim 50$ GeV this effect, not taken into account in our
computations, may be of importance.

\section{Uncertainties}

\begin{figure}
\begin{center}
\includegraphics[width=12.5cm,clip=,bbllx=0pt,bblly=30pt,bburx=560,bbury=545pt]{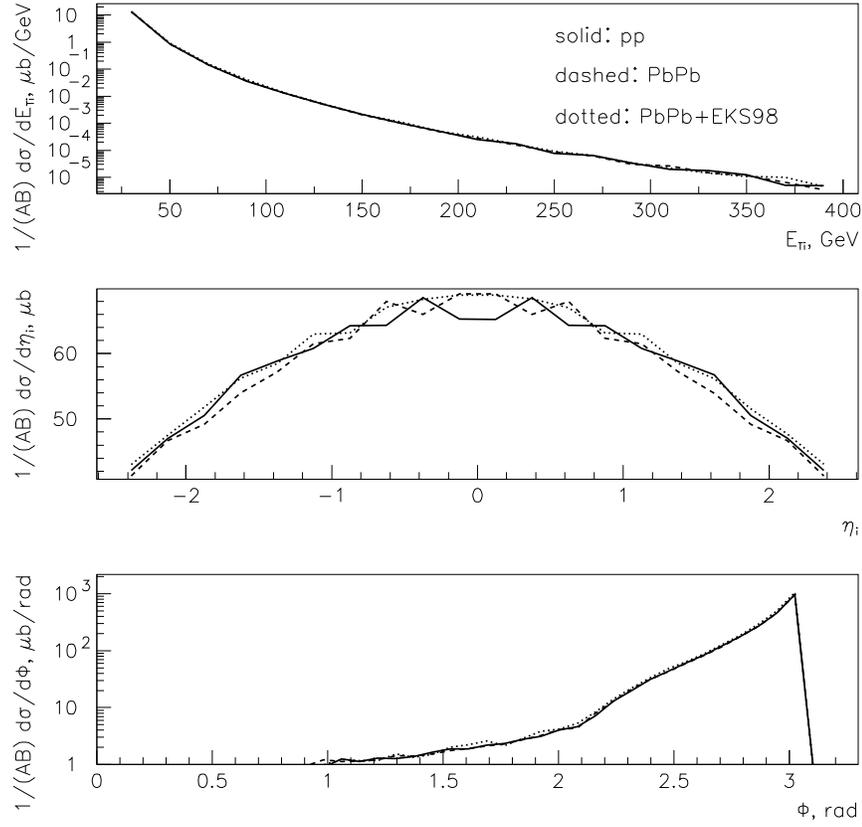}
\caption{Isospin and nuclear pdf dependence of jet cross sections (pp results:
solid lines; PbPb results without modification of nucleon pdf inside
nuclei: dashed lines; PbPb results with EKS98 modification of nucleon
pdf
inside nuclei: dotted lines) versus
transverse energy of the jet (for $|\eta_i|<2.5$, upper plot) and
pseudorapidity of the jet (for $E_{Ti}> 20$ GeV, middle plot), and
dijet cross sections (lower plot)
versus angle between the two hardest jets for $E_{T1}>20$ GeV,
$E_{T2}>15$ GeV and $|\eta_1|,|\eta_2|<2.5$, for collisions at 5.5 TeV.
Unless otherwise stated default options are
used, see text.}
\label{abfig1}
\end{center}
\end{figure}

Uncertainties on the renormalization/factorization scale, on the
jet reconstruction algorithm and on nucleon pdf, have been discussed
in the mentioned Section on pA collisions and show very similar features
in the AB case, so we will discuss them no longer. Here we will focus,
see Fig. \ref{abfig1},
on isospin effects (obtained from the
comparison of pp and PbPb without any modifications of nucleon pdf
inside nuclei at the same energy per nucleon, 5.5 TeV) and on
the effect of modifications of nucleon pdf inside nuclei,
estimated by using EKS98 \cite{eks98} nuclear corrections.

On the transverse momentum distributions isospin effects are
negligible, while effects of EKS98 result in a $\sim 3$~\% increase. On
the pseudorapidity distributions, isospin effects apparently tend to
fill a small dip at $\eta\simeq 0$ present in the pp distribution,
while EKS98 results in some increase, but nevertheless
effects never go beyond 5~\% and are not very significant when statistical
errors are considered. On the dijet angular distributions, isospin
effects are negligible while EKS98 produces an increase of order 10~\%
at maximum.

\section{Results}

\begin{figure}
\begin{center}
\includegraphics[width=16.0cm,bbllx=0pt,bblly=30pt,bburx=555,bbury=535pt]{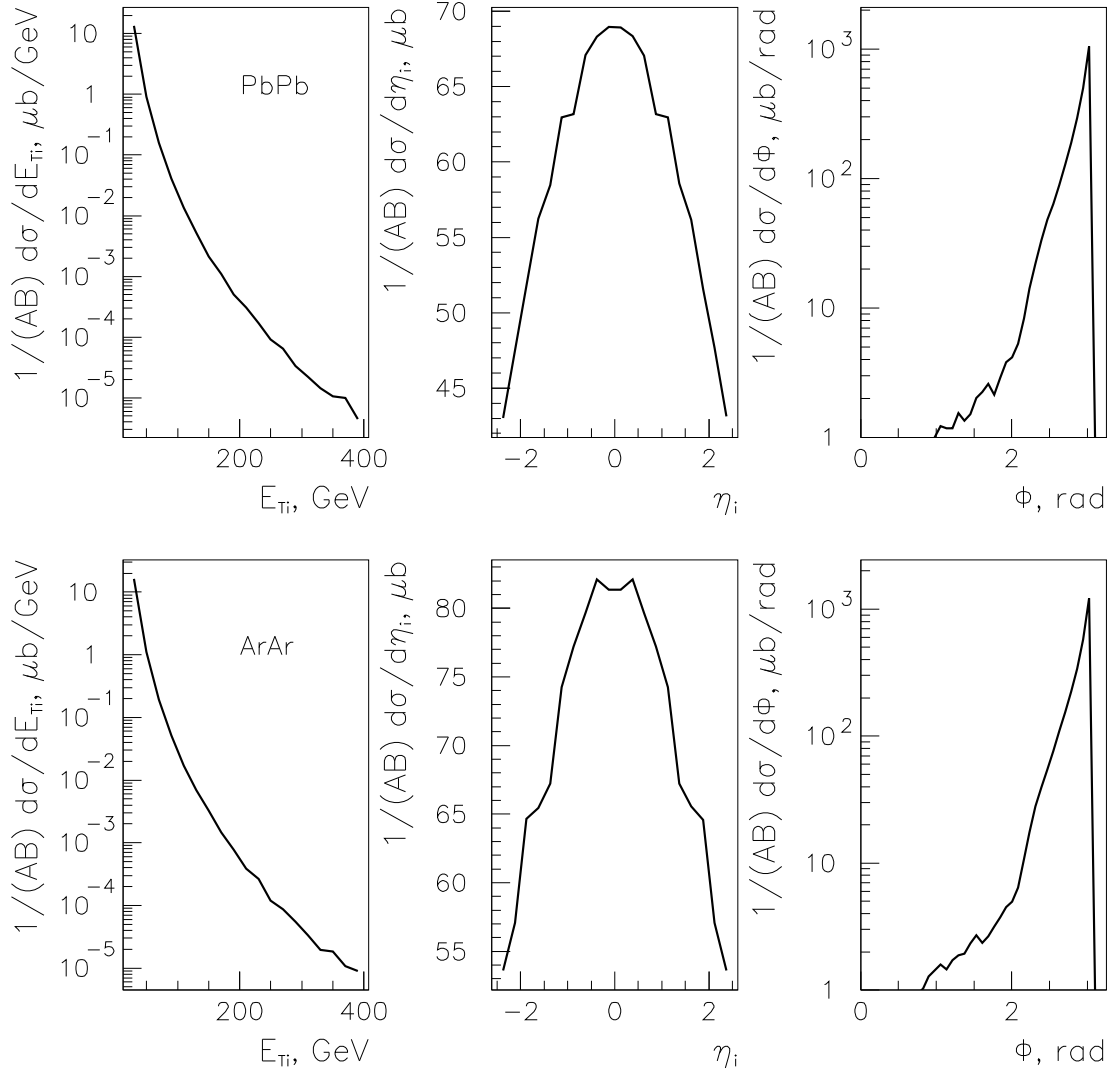}
\caption{Jet cross sections
versus
transverse energy of the jet (for $|\eta_i|<2.5$, plots on the left) and
pseudorapidity of the jet (for $E_{Ti}> 20$ GeV, plots in the middle),
and dijet cross sections
versus angle between the two hardest jets for $E_{T1}>20$ GeV,
$E_{T2}>15$ GeV and $|\eta_1|,|\eta_2|<2.5$ (plots on the right), for PbPb
collisions at 5.5 TeV (upper plots) and ArAr collisions at 6.3 TeV
(lower plots). Default options are
used, see text.}
\label{abfig2-4}
\end{center}
\end{figure}

In Table 1 the number of expected events with at least one jet with a
given $E_{Ti}> 20$ GeV and $|\eta_i|< 2.5$ (or with two jets $(1,2)$
with $E_{T1}>20$ GeV, $E_{T2}>15$ GeV and $|\eta_{1,2}|<2.5$
for the dijet $\phi$-distributions), per $\mu$b and pair of
colliding nucleons, is shown for different collisions and possible
luminosities.
From this Table and using the Figures it is possible to know the number of
expected events with a given kinematical variable. For example, examining the
solid line in Fig.
\ref{abfig2-4} (upper-left) one can expect, within the
pseudorapidity region we have considered, the following number of
jets per month in PbPb collisions at 5.5 TeV with a luminosity of $5
\cdot 10^{26}$ cm$^{-2}$s$^{-1}$: $2.2\cdot
10^{7}$ jets  with $E_{Ti}\sim 50$ GeV (corresponding to a cross section
of 1 $\mu$b/(AB)), and $2.2\cdot 10^3$ jets with $E_{Ti}\sim 250$ GeV
(corresponding to a cross section of $10^{-4}$ $\mu$b/(AB)).

A detailed study of jet quenching \cite{quench,tomo}
and of associated characteristics as jet
profiles, which should be sensitive to radiation from the jet
\cite{profile}, should be feasible with
samples of $\sim 10^3$ jets. 
Looking at the results given in Fig. \ref{abfig2-4}, it
becomes evident that, from a theoretical point of view, 
the study of such samples should be possible up to a transverse energy
$E_{Ti}\sim 275$ GeV
with a run of 1 month at the considered luminosity: indeed, from Table 1,
$10^3$ jets for PbPb would correspond to a cross section of $4.5\cdot
10^{-5}$ $\mu$b/(AB), which in Fig. \ref{abfig2-4} (upper-left) cuts the
curve at $E_{Ti}\sim 275$ GeV.

The centrality
dependence of the observables has not been examined
due to our poor knowledge of the
centrality behaviour of the modification of nucleon pdf inside nuclei;
if this behaviour becomes clear in future experiments at eA colliders
\cite{eacoll}, such study would become very useful \cite{igor}.
In any case, a variation of
nuclear sizes should allow a systematic study of the
dependence of jet spectra on the size and energy density of the produced
plasma.

\begin{table}[t]
\begin{center}
Table 1: Luminosities and expected number of events
with at least one jet with a given $E_{Ti}> 20$ GeV and $|\eta_i|< 2.5$
(or with two jets $(1,2)$ with $E_{T1}>20$ GeV, $E_{T2}>15$ GeV and
$|\eta_{1,2}|<2.5$ for the dijet $\phi$-distributions),
per $\mu$b/(AB)
in one
month ($10^6$ s), for
different collisions.
\vskip0.2cm
\begin{tabular}{|c|c|c|c|}
\hline
Collision & $E_{cm}$ per nucleon (TeV) & $\cal{L}$
(cm$^{-2}$s$^{-1}$) & Number of events per month per $\mu$b/(AB) \\
\hline
ArAr & 6.3 & $10^{29}$ & $1.6\cdot 10^8$ \\
\hline
ArAr & 6.3 & $3\cdot 10^{27}$ & $4.8\cdot 10^6$ \\
\hline
PbPb & 5.5 & $5\cdot 10^{26}$ & $2.2\cdot 10^7$ \\
\hline
\end{tabular}
\end{center}
\end{table}

\vskip 1cm
\noindent

\section*{ACKNOWLEDGEMENTS}
N. A. thanks CERN Theory Division, Departamento de F\'{\i}sica de
Part\'{\i}culas at Universidade de Santiago de Compostela, Department of
Physics at University of Jyv\"askyl\"a, Helsinki Institute of
Physics and Physics Department at BNL, for
kind hospitality during stays in which parts of this work have been
developed; he also acknowledges financial
support by CICYT of Spain under contract
AEN99-0589-C02 and by Universidad de C\'ordoba. A. A. thanks
the Max-Planck-Institut f\"ur Kernphysik of Heidelberg for permission to use
its
computing
facilities; his work is partially funded by
the European Commission IHP program under contract HPRN-CT-2000-00130.

\end{document}